\documentstyle[prl,aps,epsfig,amsmath,multicol]{revtex}

\newcommand{\mc}{\multicolumn}  
\newcommand{\lsim}{\mathrel{\mathop{\kern 0pt \rlap
  {\raise.2ex\hbox{$<$}}}
  \lower.9ex\hbox{\kern-.190em $\sim$}}}
\newcommand{\gsim}{\mathrel{\mathop{\kern 0pt \rlap
  {\raise.2ex\hbox{$>$}}}
  \lower.9ex\hbox{\kern-.190em $\sim$}}}

\title{Asymmetries in the Non--Mesonic Weak Decay of Polarized $\Lambda$--Hypernuclei}

\author{W. M. Alberico$^1$, G. Garbarino$^1$, A. Parre\~{n}o$^2$ and A. Ramos$^2$}

\address{$^1$Dipartimento di Fisica Teorica, Universit\`a di Torino
and INFN, Sezione di Torino, I--10125 Torino, Italy}

\address{$^2$Departament d'Estructura i Constituents de la Mat\`{e}ria,
Universitat de Barcelona, E--08028 Barcelona, Spain} 

\date{\today}

\begin{document}
\draft
\maketitle

\begin{abstract}
The non--mesonic weak decay of {\it polarized} 
$\Lambda$--hypernuclei is studied 
for the first time by taking into account, with a Monte Carlo intranuclear
cascade code, the nucleon final state interactions.
A one--meson--exchange model is employed
to describe the ${\vec \Lambda} N\to nN$ processes in a finite nucleus framework.
The relationship between the intrinsic $\Lambda$ asymmetry parameter
$a_\Lambda$ and the asymmetry $a^{\rm M}_\Lambda$ accessible in experiments
is discussed. A strong dependence of $a^{\rm M}_\Lambda$ on nucleon final state interactions 
and detection threshold is obtained. Our results for $a^{\rm M}_\Lambda$ are consistent with
$^{11}_\Lambda {\vec {\rm B}}$ and $^{12}_\Lambda{\vec {\rm C}}$ data but disagree with 
observations in $^5_\Lambda{\vec {\rm H}}{\rm e}$. 
\end{abstract}


\begin{multicols}{2}


The physics of the weak decay of hypernuclei has experienced a recent important 
development. Due to theoretical \cite{Ok99,Os01,Pa02,It02,ours,PBH04} 
and experimental \cite{Outa,kim,Ok} progress,
we are now towards a solution of the long standing puzzle \cite{Al02}
on the ratio, $\Gamma_n/\Gamma_p$, between the non--mesonic weak decay (NMWD) rates
for the processes
$\Lambda n\to nn$ and $\Lambda p\to np$. This has been possible mainly thanks to
the study of nucleon coincidence observables \cite{ours,Outa}. According to the
analysis of KEK data \cite{Outa} made in Refs.~\cite{ours,daph}, 
the $\Gamma_n/\Gamma_p$ ratio for both $^5_\Lambda$He and 
$^{12}_\Lambda$C is around $0.3\div0.4$, in agreement with recent 
pure theoretical estimates \cite{Ok99,Os01,Pa02,It02}. Confirmations
of these results are awaited from forthcoming experiments at DA$\Phi$NE 
\cite{finuda} and J--PARC \cite{jparc}.

Despite this recent progress, the reaction mechanism for the
hypernuclear NMWD is not fully understood. Indeed,
an intriguing problem, of more recent origin, is open: it concerns 
the asymmetry of the angular emission of NMWD protons from polarized hypernuclei.
This asymmetry is due to the interference between parity--violating
and parity--conserving ${\vec \Lambda} p\to np$ transition amplitudes \cite{Ba90}. 
The study of the asymmetric emission of protons from polarized hypernuclei
is supposed to provide information on the spin--parity
structure of the $\Lambda N\to nN$ process and hence 
new constraints on the dynamics of the non--mesonic decay.

The intensity of protons emitted in ${\vec \Lambda} p\to np$ decays
along a direction forming an angle $\theta$ with the polarization axis
is given by (for details see Ref.~\cite{Ra92}):
\begin{equation}
\label{int-w}
I(\theta)=I_0\left[1+\mathcal{A}(\theta)\right] ,
\hskip 2mm {\mathcal{A}} (\theta) = P_y\, A_y\, \cos \theta , 
\end{equation}
where $P_y$ is the hypernuclear polarization and $A_y$ the hypernuclear asymmetry
parameter. Moreover, $I_0$ is the (isotropic) intensity for an unpolarized 
hypernucleus, which we normalize as the total number of primary protons 
produced per NMWD, $I_0=1/(1+\Gamma_n/\Gamma_p)$.
In the shell model weak--coupling scheme, angular momentum algebra expresses
the polarization of the $\Lambda$ spin, $p_\Lambda$, in terms of $P_y$:
$p_\Lambda=P_y$ if $J=J_C+1/2$ and $p_\Lambda=-P_y\, J/(J+1)$ if $J=J_C-1/2$,
$J$ ($J_C$) being the hypernucleus (nuclear core) total spin.
By introducing the intrinsic $\Lambda$ asymmetry parameter,
$a_{\Lambda}=A_y$ if $J=J_C+1/2$ and $a_{\Lambda}=-A_y(J+1)/J$ if $J=J_C-1/2$,
one obtains:
${\mathcal{A}}(\theta)=p_{\Lambda}\, a_{\Lambda}\, {\rm cos}\, \theta$. 
In the hypothesis that the weak--coupling scheme provides a realistic 
description of the hypernuclear structure, $a_{\Lambda}$ can be interpreted 
as the intrinsic $\Lambda$ asymmetry parameter for the
elementary process ${\vec \Lambda} p\to np$ taking place inside the hypernucleus.
This scheme is known to be a good approximation for 
describing the ground state of $\Lambda$--hypernuclei and previous
calculations \cite{Pa02,Ra92} have proved that, thanks to the 
large momentum transfer, the non--mesonic decay 
is not much sensitive to nuclear structure details. 

Nucleon final state interactions (FSI), subsequent to the NMWD, 
are expected to modify the weak decay intensity of Eq.~(\ref{int-w}). Experimentally,
one has access to a proton intensity $I^{\rm M}(\theta)$ which is generally assumed 
to have the same $\theta$--dependence as $I(\theta)$:
\begin{equation}
\label{int-exp}
I^{\rm M}(\theta)=I^{\rm M}_0[1+p_\Lambda\, a^{\rm M}_\Lambda \cos \theta] .
\end{equation}
Then, the observable asymmetry $a^{\rm M}_\Lambda$ is determined as:
\begin{equation}
\label{a-exp}
a^{\rm M}_\Lambda =\frac{1}{p_\Lambda}\,
\frac{I^{\rm M}(0^{\circ})-I^{\rm M}(180^{\circ})}
{I^{\rm M}(0^{\circ})+I^{\rm M}(180^{\circ})} . 
\end{equation}
Concerning the determination, from data, of the intrinsic 
$\Lambda$ asymmetry parameter $a_\Lambda$, 
it is important to stress the following two questions originated
by nucleon FSI: i) one should demonstrate (experimentally and/or theoretically) that 
the angular dependence of $I^{\rm M}(\theta)$ employed in experimental analyses
is realistic; ii) if this is verified, one should investigate the relationship 
between $a_\Lambda$ and $a^{\rm M}_\Lambda$, since 
$a^{\rm M}_\Lambda$ is expected to depend on experimental conditions such as 
the proton detection threshold and the considered hypernucleus. 

The $n(\pi^+,K^+)\Lambda$ reaction is able to produce
$\Lambda$ hypernuclear states with a sizeable amount of spin--polarization \cite{Ba89}
preferentially aligned along the axis normal to the reaction plane.
Until now, four KEK experiments measured the proton asymmetric emission 
from polarized $\Lambda$--hypernuclei.
The 1992 KEK--E160 experiment \cite{Aj92}, which studied $p$-shell hypernuclei, 
suffered from large uncertainties: only
poor statistics and energy resolution could be used; moreover, the 
values of the $\Lambda$ polarization $p_\Lambda$ 
needed to determine the asymmetry $a^{\rm M}_\Lambda$, 
had to be evaluated theoretically. More recently, $a^{\rm M}_\Lambda$ 
was measured by KEK--E278 \cite{Aj00} for the 
decay of $^5_\Lambda{\vec {\rm H}{\rm e}}$. The values of $p_\Lambda$ used to 
obtain $a^{\rm M}_\Lambda$ were determined by observing the asymmetry, 
${\mathcal{A}}^{\pi^-}=p_\Lambda\, a^{\pi^-}_\Lambda$,
in the emission of negative pions in the $^5_\Lambda{\vec {\rm H}{\rm e}}$ mesonic decay, 
after assuming \cite{Mo94} $a^{\pi^-}_\Lambda$
to be equal to the value for the free $\Lambda \to \pi^- p$ 
decay, $a^{\pi^-}_\Lambda=-0.642\pm 0.013$. 
A similar measurement of $p_\Lambda$ is very difficult, instead, for 
$p$--shell hypernuclei due to their small branching ratio
and expected asymmetry ${\mathcal{A}}^{\pi^-}$ for the mesonic decay;
even the recent and more accurate experiment KEK--E508 \cite{Ma04} had to
resort to theoretical estimates \cite{MoIt} for the $\Lambda$ polarization in 
$^{12}_\Lambda{\vec {\rm C}}$ and $^{11}_\Lambda{\vec {\rm B}}$.  
Recently, $a^{\rm M}_\Lambda$ was measured again for $^5_\Lambda$He,
by KEK--E462 \cite{Ma04}, but with improved statistics.

In Table \ref{other-res} we report the results for $a^{\rm M}_\Lambda$ obtained by 
the above mentioned experiments, together with recent theoretical estimates
for $a_\Lambda$. While theoretical models predict negative $a_\Lambda$ values
\cite{note}, with a moderate dependence on the hypernucleus,
the experiments seem to favor negative values
for $a^{\rm M}_\Lambda(^{12}_\Lambda{\vec {\rm C}})$ but
positive values for $a^{\rm M}_\Lambda(^5_\Lambda{\vec {\rm H}{\rm e}})$.

Concerning the above comparison between theory and experiment,
it is important to stress that, while one predicts
$a_\Lambda(^5_\Lambda{\vec {\rm H}{\rm e}})\simeq a_\Lambda(^{12}_\Lambda{\vec {\rm C}})$,
there is no known reason to expect this approximate equality to be valid
for $a^{\rm M}_\Lambda$. Indeed, the relationship between $I(\theta)$ of Eq.~(\ref{int-w}) and 
$I^{\rm M}(\theta)$ of Eq.~(\ref{int-exp}) can be strongly affected by FSI
of the emitted protons: this fact prevents establishing a direct
relation between $a_\Lambda$ and $a^{\rm M}_\Lambda$ and to make
a direct comparison among results for these quantities.
In order to overcome this problem, in the present work we evaluate the effects of the 
nucleon FSI on the NMWD of $^5_\Lambda{\vec {\rm H}}{\rm e}$,
$^{11}_\Lambda{\vec {\rm B}}$ and $^{12}_\Lambda{\vec {\rm C}}$ 
and we perform the first theoretical estimate of $a^{\rm M}_\Lambda$. 

The $\Lambda N\to nN$ weak transition is described with the one--meson--exchange
potential of Ref.~\cite{Pa02}, which accounts for the exchange of 
$\pi$, $\rho$, $K$, $K^*$, $\omega$ and $\eta$ mesons and well reproduces
the new $\Gamma_n/\Gamma_p$ ratios extracted from KEK data \cite{Outa}
via the weak--interaction--model independent analysis of Refs.~\cite{ours,daph}.
The strong final state interactions acting
between the weak decay nucleons are taken into account through a 
scattering $nN$ wave function from the Lippmann--Schwinger equation obtained with
the Nijmegen Soft--Core NSC97 (versions ``a'' and ``f'') potentials \cite{nsc}. 
The two--nucleon stimulated
process $\Lambda NN\to nNN$ \cite{Al02,Al91} is safely neglected in our analysis.
The fraction of protons from two--nucleon induced decays which escapes from the nucleus
with an energy above the typical detection threshold is predicted \cite{ours} 
to be small with respect to the fraction originating from $\Lambda N\to nN$.
The propagation of primary (i.e., weak decay) and secondary nucleons
(due to FSI) inside the residual nucleus is simulated with the Monte Carlo code of
Ref.~\cite{Ra97}. 

\begin{figure}[htb]
\begin{center}
\mbox{\epsfig{file=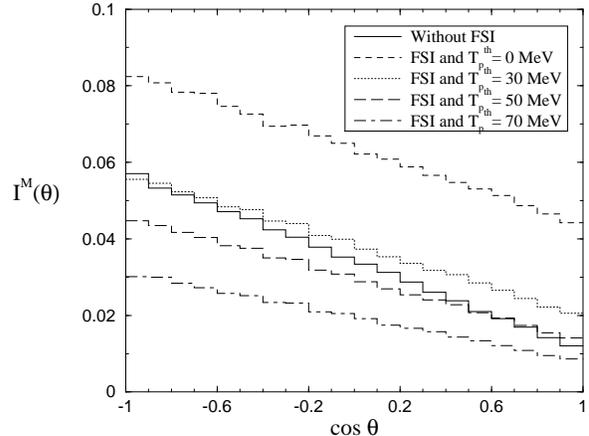,width=.45\textwidth}}
\vskip -3mm
\caption{Angular intensity of protons emitted
per NMWD of $^5_\Lambda{\vec {\rm H}{\rm e}}$.
See text for details.}
\label{he}
\end{center}
\end{figure}
\begin{figure}
\begin{center}
\mbox{\epsfig{file=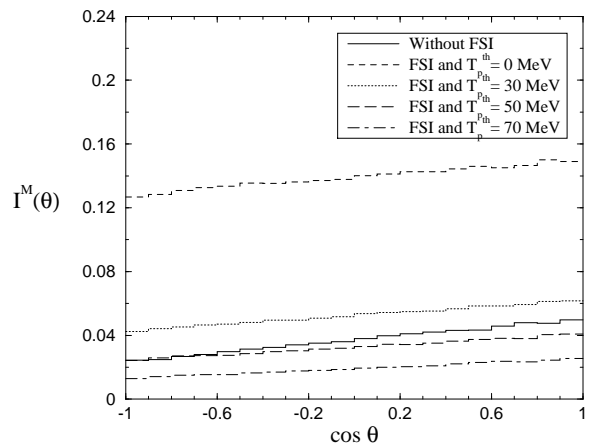,width=.45\textwidth}}
\vskip -3mm
\caption{Same of Fig.~\ref{he} for $^{12}_\Lambda{\vec {\rm C}}$. See text for details.}
\label{c}
\end{center}
\end{figure}

In Fig.~\ref{he} (\ref{c}) we show the proton intensity obtained for 
the non--mesonic decay of $^5_\Lambda{\vec {\rm H}}{\rm e}$ 
($^{12}_\Lambda{\vec {\rm C}}$) using the
full one--meson--exchange model with the NSC97f potential
(the NSC97a potential predicts very similar results). 
We note that the hypernuclear polarization has 
been taken to be $P_y=1$ in these figures, so that the hypernuclear
asymmetry parameter $A_y$ can be directly extracted from 
the values of the weak decay intensity at $\theta=0^\circ$ and $\theta=180^\circ$.
The continuous histograms correspond to the intensity $I(\theta)$ of primary protons 
[Eq.~(\ref{int-w})]. The inclusion of the nucleon FSI strongly modifies the spectra.
With vanishing kinetic energy detection threshold,
$T^{\rm th}_p$, the intensities are strongly enhanced,
especially for $^{12}_\Lambda{\vec {\rm C}}$. For $T^{\rm th}_p=30$
or $50$ MeV, the spectra are closer to $I(\theta)$, 
although with a different slope, reflecting the
fact that FSI are responsible for a substantial fraction of outgoing protons
with energy below these thresholds. A further reduction of $I^{\rm M}(\theta)$ 
is observed for $T^{\rm th}_p=70$ MeV.  
\vskip -3mm

It is evident from Figs.~\ref{he} and \ref{c} that the simulated intensities turn
out to be well fitted by the linear law in $ \cos \theta$ of 
Eq.~(\ref{int-exp}).
We can thus estimate $a^{\rm M}_\Lambda$ by using Eq.~(\ref{a-exp}) with $p_\Lambda=1$ for
$^5_\Lambda{\vec {\rm H}}{\rm e}$, $p_\Lambda=-1/2$ for $^{12}_\Lambda{\vec {\rm C}}$ and
$p_\Lambda=-5/7$ for $^{11}_\Lambda{\vec {\rm B}}$.
To do this, $I^{\rm M}(0^\circ)$ ($I^{\rm M}(180^\circ)$) is evaluated numerically as the
proton intensity in the bin with $\cos \theta\in [0.9,1]$ ($\in [-1,-0.9]$).
In Table~\ref{results} (\ref{results-p}) we show our
predictions for $I_0$, $I^{\rm M}_0$, $a_\Lambda$ and $a^{\rm M}_\Lambda$ for
$^5_\Lambda {\vec {\rm H}}{\rm e}$ ($^{11}_\Lambda {\vec {\rm B}}$
and $^{12}_\Lambda {\vec {\rm C}}$). They refer to the one--pion--exchange (OPE)
and the full one--meson--exchange (OME) models, both using the NSC97f potential. 
As a result of the nucleon FSI, $|a_\Lambda|\gsim |a^{\rm M}_\Lambda|$ for any
value of the proton threshold: when $T^{\rm th}_p=0$,
$a_\Lambda/a^{\rm M}_\Lambda\simeq 2$ for $^5_\Lambda {\vec {\rm H}}{\rm e}$
and $a_\Lambda/a^{\rm M}_\Lambda\simeq 4$ for $^{11}_\Lambda {\vec {\rm B}}$
and $^{12}_\Lambda {\vec {\rm C}}$; $|a^{\rm M}_\Lambda|$ increases
with $T^{\rm th}_p$ and $a_\Lambda/a^{\rm M}_\Lambda\approx 1$
for $T^{\rm th}_p=70$ MeV in all cases.

In Tables~\ref{results} and \ref{results-p} our results are compared
with the preliminary KEK data of Ref.~\cite{Ma04}, which correspond to a 
proton detection threshold varying (from event to event) between 
$30$ and $50$ MeV. For these conditions, we obtain OME asymmetries $a^{\rm M}_\Lambda$ 
rather independent of the hypernucleus and in the
range $-0.55\div -0.37$. The $a^{\rm M}_\Lambda$ values are smaller in size
than the corresponding asymmetries before FSI effects, $a_\Lambda$, by 25 to 50\%.
It is evident that our OME results are in agreement  with the $^{12}_\Lambda{\vec {\rm C}}$ datum,
barely compatible with the $^{11}_\Lambda{\vec {\rm B}}$ datum and inconsistent
with the $^5_\Lambda{\vec {\rm H}}{\rm e}$ datum.
One also sees that the OPE asymmetries are systematically smaller, 
though less realistic from the theoretical point of view, than the OME ones.

In view of the above large discrepancy, we have proved, numerically, 
that positive $a^{\rm M}_\Lambda$ 
values ---such as the ones measured at KEK for $^5_\Lambda{\vec {\rm H}}{\rm e}$--- can be 
obtained only if positive values for the intrinsic asymmetry $a_\Lambda$ are enforced
in the weak decay intensity $I(\theta)$ of Eq.~(\ref{int-w}): indeed,
$a_\Lambda$ and $a^{\rm M}_\Lambda$ always have the same sign.
However, unless there are large SU(3) violations in the coupling constants,
it seems unlikely that the meson--exchange models give rise
to a positive or vanishing value of the intrinsic $\Lambda$ asymmetry. 
Indeed, we have analyzed the origin of the large and negative asymmetry parameter
in the one--meson--exchange model of Ref.~\cite{Pa02}, by
calculating the two--body $\Lambda N(^{2S+1}L_J) \to n N(^{2S'+1}L'_J)$
amplitudes $a,b,c,d,e,f$ for $^5_\Lambda {\vec {\rm H}{\rm e}}$, 
and determining the intrinsic asymmetry through the following relation \cite{Na99}:
\begin{equation}
\label{asim}
a_\Lambda=\displaystyle\frac{
2\sqrt{3}\; {\rm Re} \left[ a e^* - b(c-\sqrt{2}d)^*/\sqrt{3} +
f (\sqrt{2}c+d)^* \right]}
{\mid a \mid^2+ \mid b \mid^2  + 3\left[ \mid c \mid^2 + \mid d \mid^2 +
\mid e \mid^2 + \mid f \mid^2 \right]} . \nonumber
\end{equation}
In a framework with real $\Lambda N$ and $nN$ wave functions,
the OPE mechanism produces a large and negative
$a_\Lambda$ value due, mainly, to an interference between a large and negative tensor
amplitude $d$ $(^3S_1 \to ^3\!\!D_1)$ and the parity violating
amplitudes $b$ $(^1S_0 \to ^3\!\!P_0)$ and
$f$ $(^3S_1 \to ^3\!\!P_1)$, which are both positive and of moderate size.
The inclusion of kaon exchange modifies this picture drastically.
Destructive interference with the pion in the tensor channel reduces
the $d$ amplitude by a factor of 4, which would lead to a sensitive decrease in the
size of $a_\Lambda$.
However, the negative $a$ $(^1S_0 \to ^1\!\!S_0)$ and $c$ $(^3S_1 \to ^3\!\!S_1)$
amplitudes become one order of magnitude larger in size. Their interference with the positive
$e$ $(^3S_1 \to ^1\!\!P_1)$ and $f$ $(^3S_1 \to ^3\!\!P_1)$ amplitudes end up 
producing a final value for $a_\Lambda$
which is even 50\% larger in size than for OPE alone.
The inclusion of the heavier mesons does not change this qualitative behavior.

Summarizing, we have seen how FSI are an important ingredient when studying the NMWD 
of polarized hypernuclei. The first relationship between
the intrinsic asymmetry $a_\Lambda$ and the observable asymmetry 
$a^{\rm M}_\Lambda$ has been established. 
Unfortunately, not even an analysis including FSI can 
explain the present experimental data.
From the theoretical point of view, we believe it unlikely that
new reaction mechanisms are responsible for the present discrepancies.
Only small and positive values of $a_\Lambda$, not predicted 
by any existing model,
could reduce $a^{\rm M}_\Lambda$ to small and positive values.

In order to avoid possible statistical fluctuations of the data,
new and/or improved experiments, better establishing
the sign and magnitude of $a^{\rm M}_{\Lambda}$ for $s$-- and $p$--shell hypernuclei 
(possibly also exploring the full angular region of the proton intensities) 
will be important to provide a guidance for a deeper 
understanding of the $\Lambda N\to nN$ process in nuclei.
The study of the inverse reaction
${\vec p} n\rightarrow p\Lambda$ \cite{Ki98} should also 
be encouraged since it could further supply richer and cleaner 
information on the lambda--nucleon weak interaction and especially on the 
$\Lambda$ spin--dependent observables \cite{Na99}. In our opinion, a closer 
collaboration among theoreticians and experimentalists (as the one experienced in the
recent analyses of the $\Gamma_n/\Gamma_p$ ratio) is also desirable to disclose the 
origin of the asymmetry puzzle.

Work partly supported by EURIDICE HPRN--CT--2002--00311, MIUR 2001024324\_007, INFN,
DGICYT BFM2002-01868 and Generalitat de Catalunya SGR2001-64.
Discussions with H. Bhang, T. Maruta, T. Nagae and H. Outa are acknowledged.



\begin{table}
\begin{center}
\caption{Theoretical and experimental determinations of 
the asymmetry parameters ($a_\Lambda$ and $a^{\rm M}_\Lambda$, respectively).
The predictions for $a_\Lambda$ have been obtained
with different weak transition potentials.}
\label{other-res}
\begin{tabular}{l|c c c|c}
\mc {1}{c|}{Ref. and Model} &
\mc {1}{c}{$^5_\Lambda {\vec {\rm H}}{\rm e}$} &
\mc {1}{c}{$^{12}_\Lambda {\vec {\rm C}}$} \\ \hline
K.~Sasaki et al. \cite{Ok99} &   & \\
$\pi+K+{\rm Direct \, Quark}$                & $-0.68$     \\ 
A.~Parre\~no et al. \cite{Pa02} &    & \\
$\pi+\rho+K+K^*+\omega+\eta$    & $-0.68$ & $-0.73$    \\ 
K. Itonaga et al. \cite{It03}   &    & \\
$\pi+K+\omega+2\pi/\rho+2\pi/\sigma$    & $-0.33$ &   \\ 
C. Barbero et al. \cite{Ba03}           &    & \\
$\pi+\rho+K+K^*+\omega+\eta$            & $-0.54$ &    \\
A. Parre\~no et al. \cite{PBH04} & &   \\
$\pi+K + {\rm contact\, terms}$ & $0.24$ & $0.21$ \\ \hline
  KEK--E160 \cite{Aj92}
 &                &  $-0.9\pm0.3$ *  \\
  KEK--E278 \cite{Aj00}    & $0.24\pm0.22$  &                  \\
  KEK--E508 (prel.) \cite{Ma04}    &                & $-0.44\pm0.32$  \\
  KEK--E462 (prel.) \cite{Ma04}   & $0.07\pm0.08$  &
\end{tabular}
\end{center}
\vskip -5mm
{\footnotesize * This result correspond to the weighted average (discussed on pag.~95 of 
Ref.~\cite{Al02}) among different $p$--shell hypernuclear data.}  
\end{table}
\vspace{-4mm}
\begin{table}
\begin{center}
\caption{Proton intensities and asymmetry parameters
for the non--mesonic weak decay of $^5_\Lambda{\vec {\rm H}{\rm e}}$.}
\label{results}
\begin{tabular}{l|c c}
\mc {1}{c|}{Model} &
\mc {1}{c}{$I^{\rm M}_0$} &
\mc {1}{c}{$a^{\rm M}_\Lambda$} \\ \hline
        OPE   &  & \\
{\small Without FSI ($I_0,a_\Lambda$)}                    & $0.92$  & $-0.25$     \\
{\small FSI and $T^{\rm th}_p=0$ MeV}   & $1.56$  & $-0.12$     \\
{\small FSI and $T^{\rm th}_p=30$ MeV}  & $0.99$  & $-0.18$    \\
{\small FSI and $T^{\rm th}_p=50$ MeV}  & $0.78$  & $-0.20$    \\
{\small FSI and $T^{\rm th}_p=70$ MeV}  & $0.52$  & $-0.20$    \\ \hline
            OME           & & \\
{\small Without FSI ($I_0,a_\Lambda$)}                    & $0.69$  & $-0.68$     \\
{\small FSI and $T^{\rm th}_p=0$ MeV}   & $1.27$  & $-0.30$    \\
{\small FSI and $T^{\rm th}_p=30$ MeV}  & $0.77$  & $-0.46$   \\
{\small FSI and $T^{\rm th}_p=50$ MeV}  & $0.59$  & $-0.52$   \\ 
{\small FSI and $T^{\rm th}_p=70$ MeV}  & $0.39$  & $-0.55$   \\ \hline
  KEK--E462 (prel.) \cite{Ma04}  &   & $0.07\pm 0.08$     \\
\end{tabular}
\end{center}
\end{table}
\vspace{-9mm}
\begin{table}
\begin{center}
\caption{Same as in Table~\protect\ref{results}
for $^{11}_\Lambda{\vec {\rm B}}$ and $^{12}_\Lambda{\vec {\rm C}}$.}
\label{results-p}
\begin{tabular}{l|c c| c c}
\mc {1}{c|}{Model} &
\mc {1}{c}{$^{11}_\Lambda{\vec {\rm B}}$} &
\mc {1}{c|}{} &
\mc {1}{c}{$^{12}_\Lambda{\vec {\rm C}}$} &
\mc {1}{c}{} \\
                  & $I^{\rm M}_0$    & $a^{\rm M}_\Lambda$   &
$I^{\rm M}_0$    &     $a^{\rm M}_\Lambda$ \\ \hline
        OPE                & & & & \\
{\small Without FSI  ($I_0,a_\Lambda$)}                    & $0.91$  & $-0.30$  & 
$0.93$ & $-0.34$   \\
{\small FSI and $T^{\rm th}_p=0$ MeV}   & $2.84$  & $-0.08$  & $3.15$ & $-0.09$   \\
{\small FSI and $T^{\rm th}_p=30$ MeV}  & $1.16$  & $-0.17$  & $1.22$ & $-0.20$  \\
{\small FSI and $T^{\rm th}_p=50$ MeV}  & $0.76$  & $-0.24$  & $0.78$ & $-0.28$  \\ 
{\small FSI and $T^{\rm th}_p=70$ MeV}  & $0.47$  & $-0.32$  & $0.46$ & $-0.38$  \\ \hline
            OME     & & & & \\
{\small Without FSI ($I_0,a_\Lambda$)}                    & $0.70$  & $-0.81$  & $0.75$ 
& $-0.73$   \\
{\small FSI and $T^{\rm th}_p=0$ MeV}   & $2.44$  & $-0.18$  & $2.78$ & $-0.16$  \\
{\small FSI and $T^{\rm th}_p=30$ MeV}  & $0.96$  & $-0.39$  & $1.05$ & $-0.37$  \\
{\small FSI and $T^{\rm th}_p=50$ MeV}  & $0.62$  & $-0.55$  & $0.65$ & $-0.51$  \\ 
{\small FSI and $T^{\rm th}_p=70$ MeV}  & $0.38$  & $-0.70$  & $0.38$ & $-0.65$  \\ \hline
  KEK--E508 (prel.) \cite{Ma04}  &   & $0.11\pm 0.44$  &  & $-0.44\pm 0.32$  \\
\end{tabular} 
\end{center}
\end{table}

\end{multicols}

\begin{references}
\bibitem{note}
The only exception is the work of Ref.~\cite{PBH04}, an
effective theory which incorporates the exchange of a pion and a kaon plus
contact terms up to order one in a momentum expansion.
Reasonable values for the low--energy coefficients are obtained when
a fit which includes $\Gamma_{\rm NM}$ and $\Gamma_n / \Gamma_p$ for
$^5_\Lambda$He, $^{11}_\Lambda$B and $^{12}_\Lambda$C,
together with $a_\Lambda (^5_\Lambda \rm{He})=0.24$ \cite{Aj00}, is performed.
\end{references}
\end{document}